\documentclass{article}

\usepackage{graphicx}
\usepackage{lscape}
\usepackage{float}
\usepackage{setspace}
\usepackage{amssymb}
\usepackage{amsmath,nicefrac}
\usepackage{fixltx2e}
\usepackage{fancyhdr}
\usepackage{rotating}
\usepackage{wrapfig}
\usepackage{textcomp}
\usepackage[small]{caption} 
\usepackage{cite}
\usepackage{subfigure}
\usepackage{color}
\usepackage[pdfpagelabels]{hyperref}
\usepackage{authblk}

\title{Attosecond Transient Absorption Spooktroscopy: a ghost imaging approach to ultrafast absorption spectroscopy}
\author[*,1,2,3]{Taran~Driver}
\author[,2,4]{Siqi~Li\thanks{Authors contributed equally}}
\author[1]{Elio~G.~Champenois}
\author[2]{Joseph~Duris}
\author[2]{Daniel~Ratner}
\author[2]{TJ~Lane}
\author[1,5,6]{Philipp~Rosenberger}
\author[7]{Andre~Al-Haddad}
\author[3]{Vitali~Averbukh}
\author[3]{Toby~Barnard}
\author[8]{Nora~Berrah}
\author[7,9]{Christoph~Bostedt}
\author[1,4,10]{Philip~H.~Bucksbaum}
\author[1,2]{Ryan~Coffee}
\author[11]{Louis~F.~DiMauro}
\author[11]{Li~Fang}
\author[3]{Douglas~Garratt}
\author[2]{Averell~Gatton}
\author[1,4]{Zhaoheng~Guo}
\author[12]{Gregor~Hartmann}
\author[13]{Daniel~Haxton}
\author[14]{Wolfram~Helml}
\author[2,4]{Zhirong~Huang}
\author[8]{Aaron~LaForge}
\author[2,3]{Andrei~Kamalov}
\author[1,5,6]{Matthias~F.~Kling}
\author[1]{Jonas~Knurr}
\author[2]{Ming-Fu~Lin}
\author[2]{Alberto~A.~Lutman}
\author[2,4]{James~P.~MacArthur}
\author[3]{Jon~P.~Marangos}
\author[2,4]{Megan~Nantel}
\author[1]{Adi~Natan}
\author[8]{Razib~Obaid}
\author[2]{Niranjan~H.~Shivaram}
\author[1]{Aviad Schori}
\author[2]{Peter~Walter}
\author[1,10]{Anna~Wang}
\author[1]{Thomas~J.~A.~Wolf}
\author[2]{Agostino~Marinelli\thanks{Author to whom correspondance should be addressed: marinelli@slac.stanford.edu}}
\author[1,2]{James~P.~Cryan\thanks{Author to whom correspondance should be addressed: jcryan@slac.stanford.edu}}

\affil[1]{Stanford PULSE Institute, SLAC National Accelerator Laboratory}
\affil[2]{SLAC National Accelerator Laboratory}
\affil[3]{The Blackett Laboratory, Department of Physics, Imperial College
London}
\affil[4]{Department of Physics, Stanford University}
\affil[5]{Max-Planck-Institut f\"ur Quantenoptik}
\affil[6]{Physik Department, Ludwig-Maximilians-Universit\"at Munich}
\affil[7]{Paul Scherrer Institut}
\affil[8]{Physics Department, University of Connecticut}
\affil[9]{Ecole Polytechnique F\'ed\'erale de Lausanne}
\affil[10]{Department of Applied Physics, Stanford University}
\affil[11]{Department of Physics, The Ohio State University}
\affil[12]{Institut f\"ur Physik und CINSaT, Universit¨at Kassel}
\affil[13]{KLA Tencor Corporation}
\affil[14]{Zentrum f\"ur Synchrotronstrahlung, Technische Universit\"at Dortmund}

\date{\today}

\begin{document}
\maketitle

\begin{abstract}
The recent demonstration of isolated attosecond pulses from an X-ray free-electron laser (XFEL) opens the possibility for probing ultrafast electron dynamics %with previously inaccessible nonlinear spectroscopic schemes.
at X-ray wavelengths.
An established experimental method for probing %\textit{femtosecond} 
ultrafast
dynamics is X-ray transient absorption spectroscopy,
where the X-ray absorption spectrum is measured %at free-electron lasers 
by scanning the central photon energy and recording the resultant photoproducts. 
The spectral bandwidth inherent to attosecond pulses is wide compared to the resonant features typically probed, which generally precludes the application of this technique in the attosecond regime.
In this paper we propose and demonstrate a new technique to conduct transient absorption spectroscopy with broad bandwidth attosecond pulses with the aid of ghost imaging, recovering sub-bandwidth resolution in photoproduct-based absorption measurements.
\end{abstract}

%\textbf{TODO}
%\begin{itemize}
    %\item signal-to-noise figure
    %\item send to collaborators
    %\item label in panel c of fig 1
    %\item be consistent with terminology
    %\item press difference between first discussion figure (why experiment worked) and second (what the limits of the technique are)
    %\item references
    %\item change SNR to mode of Siqi's distribution
%    \item at the end of discussion, add a few sentences to summarize all the effects, and advise on experiment design.
    %\item check discussion of Dan's narrower feature is correct
    %\item push the fact that traditional way of doing it only changes first moment
    %\item mention prefer higher res to more variation - reconstruction breaks down faster with reduced resolution vs reduced variation
    %\item why the measured second moment of the data is so much larger than expected at 532 eV
%\end{itemize}

\section*{Introduction}
The rearrangement of electrons is the first step in all chemical reactions.
The ability to produce pulses with a time duration shorter than a femtosecond has enabled the probing of electron dynamics on its natural timescale~\cite{krausz_attosecond_2009,leone_what_2014,lepine_attosecond_2014,nisoli_attosecond_2017}.
This attosecond revolution has been led by the development of sources based on high harmonic generation~(HHG)~\cite{chini_generation_2014}.
These technological developments have enabled the study of autoionization~\cite{wang2010attosecond}, Auger decay~\cite{drescher_time-resolved_2002}, and charge migration~\cite{calegari_ultrafast_2014} in the time-domain.
The extension of attosecond pulses to soft X-ray wavelengths should enable the study of coherent electronic phenomena with atomic site specificity.
One particular class of attosecond spectroscopy commonly used with attosecond sources is attosecond transient absorption spectroscopy~(ATAS)~\cite{goulielmakis_attosecond_2007,sansone_electron_2012,attar_femtosecond_2017,pertot_time-resolved_2017}.
ATAS measures the spectral response of a sample after sequential interaction with a pump pulse and a broadband attosecond probe pulse.
Using photon energies in the soft X-ray regime, ATAS can probe resonant transitions between inner valence or core electrons and unoccupied states in the valence shell~\cite{bressler_ultrafast_2004}. 
The spatial localization of the core orbitals means excitation of electrons from these orbitals to valence electronic states provides an atomic-site specific probe of transient valence electronic structure.
The recent demonstration of isolated attosecond pulses from an X-ray free-electron laser (XFEL), with photon energies tunable across the soft X-ray regime and spectral brightness six orders of magnitude greater than HHG sources~\cite{duris2019}, enables numerous previously unfeasible attosecond measurements.
However, the implementation of ATAS with an attosecond XFEL source presents a number of challenges.
Here, we introduce and demonstrate a new experimental implementation to perform ATAS at an XFEL.

ATAS is an extension of transient X-ray absorption spectroscopy~(TAS), which is itself a time domain implementation of X-ray absorption spectroscopy.
Traditional X-ray absorption spectroscopy of a target is performed by scanning the central photon energy of a narrow linewidth X-ray source.
For each photon energy the total number of photons absorbed by the target is measured, either by directly measuring a depletion in the transmitted photons~\cite{bressler_ultrafast_2004} or by measuring the total ion or electron~(photoproducts) yield produced from the target following interaction with the incident light.
The latter serves as a direct indicator of the number of absorbed photons. 
In TAS, a pump laser pulse first creates an excited state in the system being probed.
The time evolution of this excited state is mapped out by measuring the X-ray absorption spectrum as a function of delay between the pump and X-ray pulses.
This technique has proved successful in measuring the ultrafast evolution of excited systems.
For example, Wolf~\textit{et al.} employed this technique at an XFEL by scanning a narrow bandwidth X-ray pulse over the near-edge features of gas phase thymine in order to observe the molecular deactivation process following ultraviolet excitation~\cite{wolf_probing_2017}.

Probing dynamics that evolve on the femtosecond or sub-femtosecond timescale requires probe pulses with a broad bandwidth (a Fourier transform limited light pulse with 0.5~fs duration has a spectral bandwidth of 3.2~eV). 
Therefore, the Fourier limit fundamentally restricts the application of traditional photoproduct-based TAS measurements on the attosecond timescale.
Simply scanning the central wavelength of a sub-femtosecond X-ray pulse will yield an X-ray absorption spectrum with poor resolution. 

To resolve this issue, ATAS was developed to measure the attosecond transient absorption spectrum of excited samples.
%ATAS experiments with HHG-based X-ray and XUV sources have found an elegant solution to this problem. 
In ATAS, transient absorption spectra are obtained by spectrally resolving the depletion in the number of probe pulse photons transmitted through the sample under analysis~\cite{beck_probing_2015}. 
The transmitted light is dispersed with a grating and measured on a spatially resolving detector.
Measurement of the transient absorption spectrum at a specific pump-probe delay involves acquiring a reference spectrum without the sample or the pump pulse, and comparing this to the absorption spectrum taken at the pump-probe delay in question.
%Reference spectra can be measured without the sample and without the pump pulse and compared to the transient absorption spectrum.
%The difference between a reference spectrum and the transient absorption spectrum yields a measurement of the target's absorption spectrum at a particular pump-probe delay.
The advantage of this technique is that the spectral resolution of the measurement is dictated by the spectrometer resolution, and not by the bandwidth of the incident pulse.
This breaks the requirement for narrow bandwidth pulses to maintain practical spectral resolution.
%ATAS using HHG-based X-ray and XUV sources is only possible because transient absorption spectra can be measured at high spectral resolution with the broad bandwidths required to produce attosecond laser pulses.
The broad success of such measurements in resolving ultrafast dynamics in molecules and solids is the topic of recent reviews~\cite{geneaux2019transient,ramasesha_real-time_2016}. %there were too many 

X-ray free electron lasers~(XFELs) are an emerging source of ultrafast soft X-ray pulses with few-femtosecond to sub-femtosecond duration ~\cite{coffee_ryan_n._development_2019}.
The tunability and unprecedented brightness of XFEL sources provides a powerful tool for the experimental investigation of ultrafast molecular dynamics~\cite{berrah2017perspective}. 
%XFELs can produce X-ray pulses with tens of gigawatts of peak power, at least one million times higher than HHG sources.
Recently, GW-scale soft X-ray isolated attosecond pulses (IAPs) were demonstrated at the Linac Coherent Light Source~(LCLS), using an implementation of the enhanced self amplified spontaneous emission~(ESASE) technique~\cite{duris2019}.
The spectral brightness of this attosecond source is six orders of magnitude greater than any tabletop HHG-based source of IAPs, facilitating non-linear spectroscopies.
The inherent spectral bandwidth of attosecond pulses greatly limits the spectral resolution of traditional photoproduct-based absorption measurements.
However, instabilities, which are inherent to XFEL operation, make implementing standard ATAS difficult. 
A photon depletion-based measurement benefits from a highly stable spectrum.
Since the number of absorbed photons is determined by taking a difference between transmitted spectrum with and without the target, variation in the spectrum of the source produces differences between reference and measurement spectra.
This adds noise to the transient absorption measurement.
The signal-to-noise of an ATAS measurement also places stringent requirements on the density and absorption cross section of the sample under analysis.
For a good quality measurement, the spectrum must be stable and the photon depletion must be sufficient to be both measurable, and significant compared to the spectral instabilities.

So while it is possible to implement a photon depletion-based measurement at an XFEL, a photoproduct-based scheme is desirable for a number of reasons.
%Detection of photoproducts is an alternative to photon-depletion spectroscopy.
Electrons or ions produced by photoemission can be detected with very high efficiency, even permitting detection of the absorption of a single X-ray photon. 
Additionally, the information content of a photoproduct-based scheme surpasses that of photon-depletion spectroscopy if the energy, mass or angular distribution of the photoproducts can also be measured.
In this work, we demonstrate how correlation techniques can be used to recover a high resolution absorption spectrum from a photoproduct-based measurement when the probing X-ray pulse has a bandwidth larger than the absorption features being measured.
The photoproduct yield can be correlated with shot-to-shot changes in the incident X-ray spectrum. 
Absorption spectra may be derived from these correlations using an algorithm that is related to so-called `ghost imaging' methods.

\subsection*{Ghost Imaging}
Classical ghost imaging is an experimental technique which can retrieve spatially resolved information about a sample using only a single pixel camera (or  ``bucket'' detector) and knowledge of the spatial structure of the illuminating source before the sample. 
A classical ghost imaging experiment often consists of a beam splitter that separates the incident wavefront into two arms.
One arm is used to analyze the wavefront with some form of pixelated detector.
For every exposure $i$, the detector records the wavefront $A_i$. 
The other arm passes through the sample under analysis, and the total transmission, $b_i$, is measured by the bucket detector. 
%In the situation where one encodes a known pattern onto the illuminating source through the use of spatial light modulators, the first path is eliminated, and this setup is called ``computational ghost imaging''. 
The coincident measurement of the two arms is repeated many times.
By correlating the shot-to-shot variation in the patterned wavefront with the measured bucket intensity that each wavefront produces, the structure of the sample can be inferred without directly detecting it, hence the term ``ghost imaging''.

The ghost imaging problem can be formulated mathematically, across $n$ different measurements with a pixelated detector of $m$ pixels, as a linear matrix multiplication:
\begin{equation}
    \mathbf{b}=\mathbf{A}\mathbf{x}.
\label{eq::GIequation}
\end{equation}
Here, $\mathbf{b}$ is a length-$n$ column vector where each element is the bucket detector reading for each measurement $b_i$, $\mathbf{A}$ is the $n \times m$ matrix of each pixelated measurement of the incident wavefront, and $\mathbf{x}$ is the length-$m$ row vector of the unknown variable to be reconstructed.
The solution to an equation of form Eqn.~\ref{eq::GIequation} has been widely studied in many research fields and has led to a variety of algorithms to invert the equation to solve for $\mathbf{x}$ when the matrix $\mathbf{A}$ is not trivially invertible, e.g. in the case of an underconstrained problem or a noisy measurement.
The ghost imaging scheme is especially useful in experiments where pixelated detection is challenging or the sample under analysis is radiation sensitive.
It has been widely demonstrated in the spatial domain with various illuminating sources, including visible light, X-rays, atoms, and electrons~\cite{erkmen10s,duarte08dt,pelliccia16rs,yu16lh,zhang18hw,khakimov16hs,schori2017x,schori2018ghost,li2018electron,klein2019x}, as well as being demonstrated in the spectral domain \cite{amiot2018supercontinuum,janassek2018ghost,kalashnikov2016infrared,scarcelli2003remote}.
Ghost imaging in the time domain has also been demonstrated \cite{ryczkowski2016gitime}, and has been proposed for X-ray pump/X-ray probe experiments with a single self-amplified spontaneous emission (SASE) pulse~\cite{ratner2019pump}, 
using the inherent stochastic nature of SASE pulses to extract time-resolved pump-probe measurements.

In this work, we exploit the natural fluctuations in the spectral profile of ESASE pulses and apply ghost imaging in the spectral domain to recover sub-bandwidth resolution photoproduct-based absorption spectra using attosecond X-ray pulses from an XFEL.
Our demonstration enables attosecond-resolution transient absorption experiments using information-rich photoproduct measurements, and with spectral resolution limited only by the properties of the photon spectrometer and the variation in the X-ray spectrum.

\section*{Experimental Demonstration}
\begin{figure}
\centering  
    \resizebox{0.8\columnwidth}{!}{\includegraphics{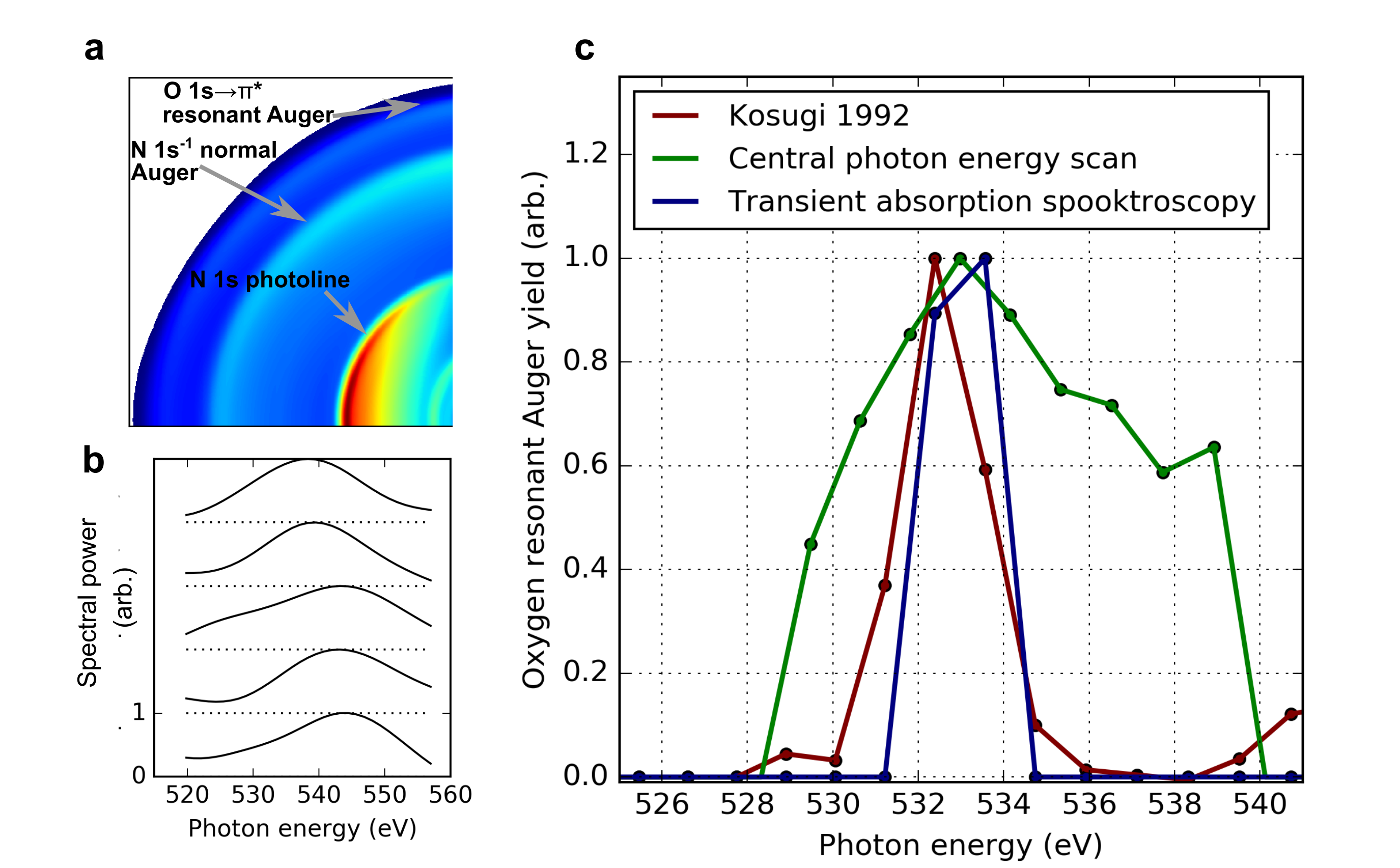}}
    \caption{a) Inverted c-VMI image showing simultaneous measurement of nitrogen 1s photoline, nitrogen normal Auger, and oxygen resonant Auger. b) Example ESASE pulses spectra measured at the nitrogen 1s photoline. The high c-VMI plate voltages result in poor spectral resolution, significantly lower than for a typical photon or electron spectrometer within its standard operating regime. As a result, the measured bandwidth is significantly higher than measured with a higher resolution spectrometer. c) Comparison of the standard method of absorption spectrum retrieval using the electron beam energy (green) \textit{vs.} absorption spooktroscopy (blue).}
    \label{fig:GAspectra6_recon}
\end{figure}

To demonstrate our technique, we have performed an experiment at the LCLS using a static absorption measurement at the oxygen K-edge of nitric oxide (NO).
The attosecond X-ray pulses used to excite the system are generated by the ESASE method~\cite{duris2019}.
%We use a single electron spectrometer to serve as the pixelated camera as well as the bucket detector.
The attosecond X-ray pulse is tuned near the $1s\rightarrow2\pi^{\ast}$ resonance of the oxygen atom in NO, at 532.7~eV~\cite{kosugi_highresolution_1992}.
At this photon energy, the attosecond ESASE pulses have a median bandwidth of $\sim5.5$~eV~\cite{duris2019}. 
Using a co-axial velocity map imaging spectrometer~(c-VMI)~\cite{li_co-axial_2018}, which projects the charged particle momentum distribution along the propagation axis of the X-rays, we measure photoelectrons with kinetic energies up to $\sim$600~eV.
This enables measurement of the resonant Auger electrons produced following relaxation of the resonant core excitation.
The inverted c-VMI image is shown in panel \textbf{a} of Fig. \ref{fig:GAspectra6_recon}.
Some functional imperfections at the high operating voltages of the c-VMI in this mode of operation contribute to the measured angular anisotropy of the normal and resonant Auger in panel \textbf{a} of Fig. \ref{fig:GAspectra6_recon}.
The energy resolution of the c-VMI spectrometer is $\sim1\%$~($\nicefrac{\Delta E}{E}$), which precludes a spectrally resolved measurement of the high energy resonant Auger electron spectrum, but is sufficient to measure the total resonant Auger yield.
This provides a direct measurement of the $1s \rightarrow 2\pi^{\ast}$ absorption cross section.

As described above, a standard measurement of the near-edge absorption spectrum involves scanning the central photon energy and measuring the yield of resultant photoproducts at each energy.
This was performed by Kosugi~\textit{et al.}~\cite{kosugi_highresolution_1992} who used narrowband synchrotron radiation to obtain a high-resolution measurement of the oxygen 1s$\rightarrow \pi^{*}$ resonance, and this data is reproduced in red in panel \textbf{c} of Fig.~\ref{fig:GAspectra6_recon}.
At the LCLS, this measurement involves scanning the easily adjustable energy of the lasing electron beam%(as performed in \cite{wolf_probing_2017})
, which changes the wavelength of X-ray radiation according to the well-known FEL resonance condition~\cite{pellegrini2016}.
Performing this measurement using the broadband attosecond ESASE pulses produces the absorption spectrum plotted in green in Fig.~\ref{fig:GAspectra6_recon}.
The measurement of this absorption feature has extremely low resolution because it effectively involves the convolution of the resonant feature with the broad bandwidth of the attosecond pulses and other photon energy jitter.
 
Using the correlation present in the data set, we can increase the resolution of the absorption measurement and obtain sub-bandwidth resolution. 
%Here, we demonstrate a measurement of the same absorption spectrum at higher, sub-bandwidth resolution using ghost imaging.
Casting the problem in the form of Eqn.~\ref{eq::GIequation}, we consider 4060 individual measurements, corresponding to the same number of single XFEL shots each at pulse energy 70~$\mu$J or higher.
The yield of resonant Auger electrons is used as the bucket detector reading, $\mathbf{b}$, and the single-shot nitrogen photoelectron spectrum is used as the pixelated detection, $\mathbf{A}$.
The unknown vector $\mathbf{x}$ now represents the target absorption spectrum.
The nitrogen 1s photoline serves as an approximate measurement for the X-ray spectrum, and is recorded by the c-VMI in coincidence with the Auger electron yield on a shot-to-shot basis, as shown in panel~\textbf{a} of Fig. \ref{fig:GAspectra6_recon}.
Some example shot-to-shot photon spectra taken from this measurement are shown in panel~\textbf{b}.
The limited resolution of the c-VMI for high energy operations results in lower than expected resolution for an X-ray photon spectrometer.
%The high operating voltages on the c-VMI plates limit the resolution of this measurement, which is much lower than expected for an X-ray photon spectrometer or electron spectrometer operating under standard parameters.
We use a standard optimization tool known as the alternating direction method of multipliers (ADMM,~\cite{boyd11pc}), to solve Eqn.~\ref{eq::GIequation}. 
This enables regularization of the retrieved solution according to expected non-negativity, sparseness, and smoothness.
We also account for a pulse energy-dependent electron background by introducing an extra term to allow the algorithm to separate the signal from the background-related contributions.
So Eqn.~\ref{eq::GIequation} becomes
\begin{equation}
    %\mathbf{b}=A\mathbf{x}+\mathbf{\epsilon}
    %\mathbf{b}=[\mathbf{A},\mathbf{P}][\mathbf{x} ;\mathbf{x_{\epsilon}}],
    \mathbf{b} = \mathbf{A} \mathbf{x} + \mathbf{P}\mathbf{x_0}
    \label{eq::GIequation_2}
\end{equation}
where $\mathbf{P}$ is the vector of single-shot pulse energies, and $\mathbf{x_0}$ is the pulse energy dependence of the background. 
The results obtained by 4060 c-VMI spectrum measurements with corresponding Auger yields are plotted in blue in panel~\textit{c} of Fig.~\ref{fig:GAspectra6_recon},
showing very good agreement with the measured high resolution absorption spectrum from Kosugi~\textit{et al.} \cite{kosugi_highresolution_1992}.
The retrieved full-width-half-maximum (FWHM) of the resonant feature~($<$4~eV) is significantly smaller than the photon bandwidth.
We note that use of a sequential least squares quadratic programming optimization to solve Eqn. \ref{eq::GIequation_2} with a simple Gaussian and single pulse energy dependent parameter recovered the correct position of the resonance, although it returned a significantly smaller width.

We performed a comparison to characterize the signal-to-noise in our experimental measurement.
For the measured spectrum and pulse energy of each of the 4060 shots, we use the retrieved values of $\mathbf{x}$ and $\mathbf{x_0}$ to calculate the predicted value which we would expect to measure in the bucket detector for that shot.
We compare this expected value to the actual measured value of the resonant Auger yield for each shot.
This is not perfectly equivalent to the signal-to-noise of the overall measurement, because there is noise in the spectral measurement as well as the bucket measurement.
Nonetheless, it serves as a useful approximation to the overall fidelity of the experimental measurement.
We find that the actual bucket measurement deviates from the expected bucket measurement by $\sim\nicefrac{2}{5}$ of the expected measurement yield, on average.
The noise in our experimental measurement has been overcome by significantly overdetermining Eqn.~\ref{eq::GIequation_2} using a large number of single shot measurements. %I meant to say overdetermining here!
To check the validity of our reconstruction, we performed one hundred separate randomizations of the order of the bucket detector and repeated the reconstruction for each iteration.
The algorithm did not retrieve the correct absorption spectrum on any iteration.
Our experimental results indicate that the variation inherent to ESASE operation provides sufficient shot-to-shot spectral differences to successfully reconstruct the sub-bandwidth absorption spectrum.

\section*{Discussion}
\begin{figure}
\resizebox{0.99\columnwidth}{!}{\includegraphics{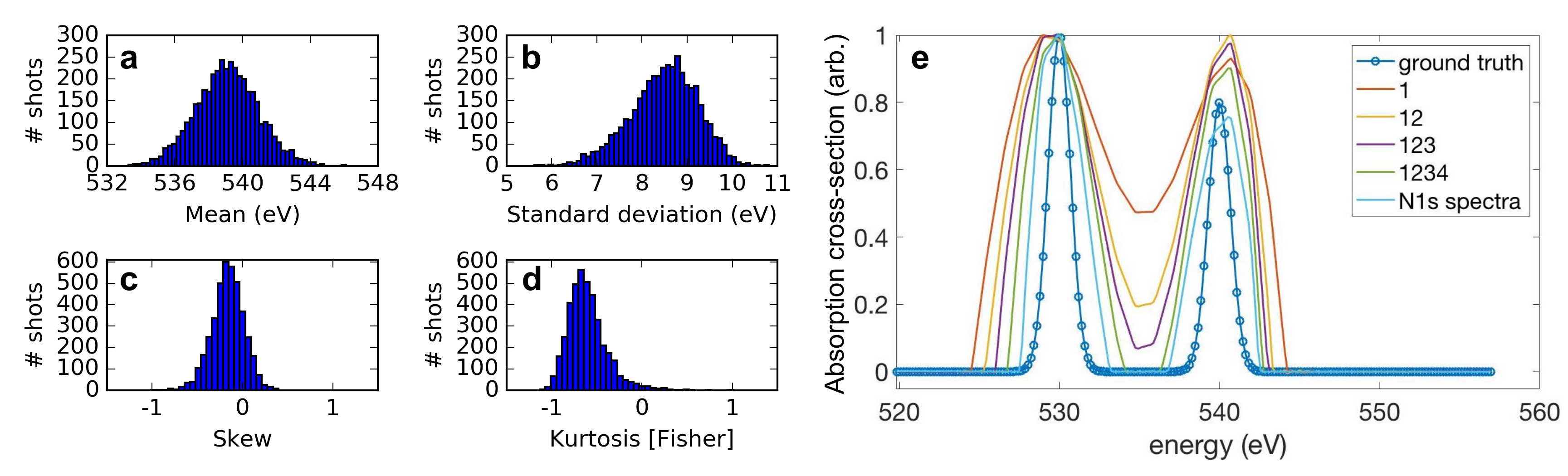}}
\caption{\noindent a)-d) Shot-to-shot variation of the four normalized central moments of the spectra used in Fig.~\ref{fig:GAspectra6_recon}. e) Performance of ADMM as the full variation of each moment is added. Using the measured variation in each of the four moments as shown in panels \textbf{a}-\textbf{d}, we simulated different sets of 4060 random spectra where some moments were allowed to vary as measured (these are indicated in the figure legend) and the variation of some moments was artificially restricted. The absorption spectra retrieved for each of these different sets of spectra is plotted against the ground truth used, showing the importance of variation in higher order moments to our technique. ``N1s spectra'' labels the actual experimental data measured by the c-VMI.}
\label{fig:moments}

\end{figure}

For our technique to work, it is critical to have sufficient variation in the measurement matrix $\mathbf{A}$, which corresponds to sufficient shot-to-shot variation in the X-ray spectrum for our measurements. 
%the X-ray spectrum, or in , and 
Moreover, the variation has to be detectable within the instrument resolution.
%\textbf{in the same figure, include single shot spectra of the VMI measurement. Then in another figure, show the four moments distribution.}
We explore the spectral variation necessary for our correlation analysis using two different metrics.

First, we explore the different modes of variation in our experimental spectrum measurements.
%according to their determined central moments.
Each of the 4060 c-VMI measured spectra (representative examples in panel~\textit{b} of Fig.~\ref{fig:GAspectra6_recon}) can be well-approximated by the Gram-Charlier expansion of a Gaussian curve with distortion given by additional moments, up to the fourth moment.
These four central moments give the mean, standard deviation, skew, and (excess/Fisher) kurtosis of a curve, respectively.
The shot-to-shot variation of these moments is plotted in panels \textbf{a}--\textbf{d} of Fig.~\ref{fig:moments}.
The measured standard deviation corresponds to a significantly larger spectral bandwidth than the $\sim5.5$~eV~\cite{duris2019} measured for ESASE pulses at these photon energies, which is a direct result of the reduced resolution of the c-VMI spectrometer when collecting high energy electrons. %at these non-standard operating voltages.
Using the moment expansion allows us to isolate the variation of each of the spectral moments to explore the dependence of the correlation method on each of these parameters.
The traditional method of scanning the central photon energy corresponds to the limit of changing only the first moment.
From these moment distributions, we construct different sets of 4060 simulated X-ray spectra to test the performance of ADMM in simulation.
First, we artificially narrow the variation in the second, third and fourth moments by a factor of two, while allowing the first moment to vary according to the measured experimental distribution.
%3489 measured with X-ray spectrometer, 4060 from VMI (at 70 uJ or higher)
We then create additional sets of simulated spectra where each higher-order moment is consecutively allowed to vary fully according to its measured distribution, sequentially increasing the contribution to the spectral variation from higher-order moments.

To test the reconstruction using these simulated data sets, we simulate a ground-truth absorption spectrum of two Gaussian peaks separated by 10~eV with root-mean-square~(rms) width of 0.7~eV. 
The simulated bucket detector reading is generated by multiplying the simulated spectra with the ground truth.
The results of the different reconstructions are shown in Fig.~\ref{fig:moments}.
% The numerical labels indicate the moments which are allowed to vary fully in the simulated spectra, and ``N1s spectra'' labels the experimental data itself.
Including higher moments in the spectra captures more spectral variation and therefore improves the resolution of the reconstruction. 
The highly similar performance of the reconstruction where all four moments may vary fully and the experimental data indicates that all the experimental variation is well captured by the four moments.
%It is worth noting that traditional measurement of scanning the central photon energy to obtain an absorption spectrum is equivalent to only varying the first moment. 
Therefore in our ghost imaging technique, we exploit higher moments of variation in the spectra to retrieve sub-bandwidth structures in the absorption spectrum. %this is a very nice way of thinking about it

\begin{figure}
\begin{centering}
\resizebox{0.7\columnwidth}{!}{\includegraphics{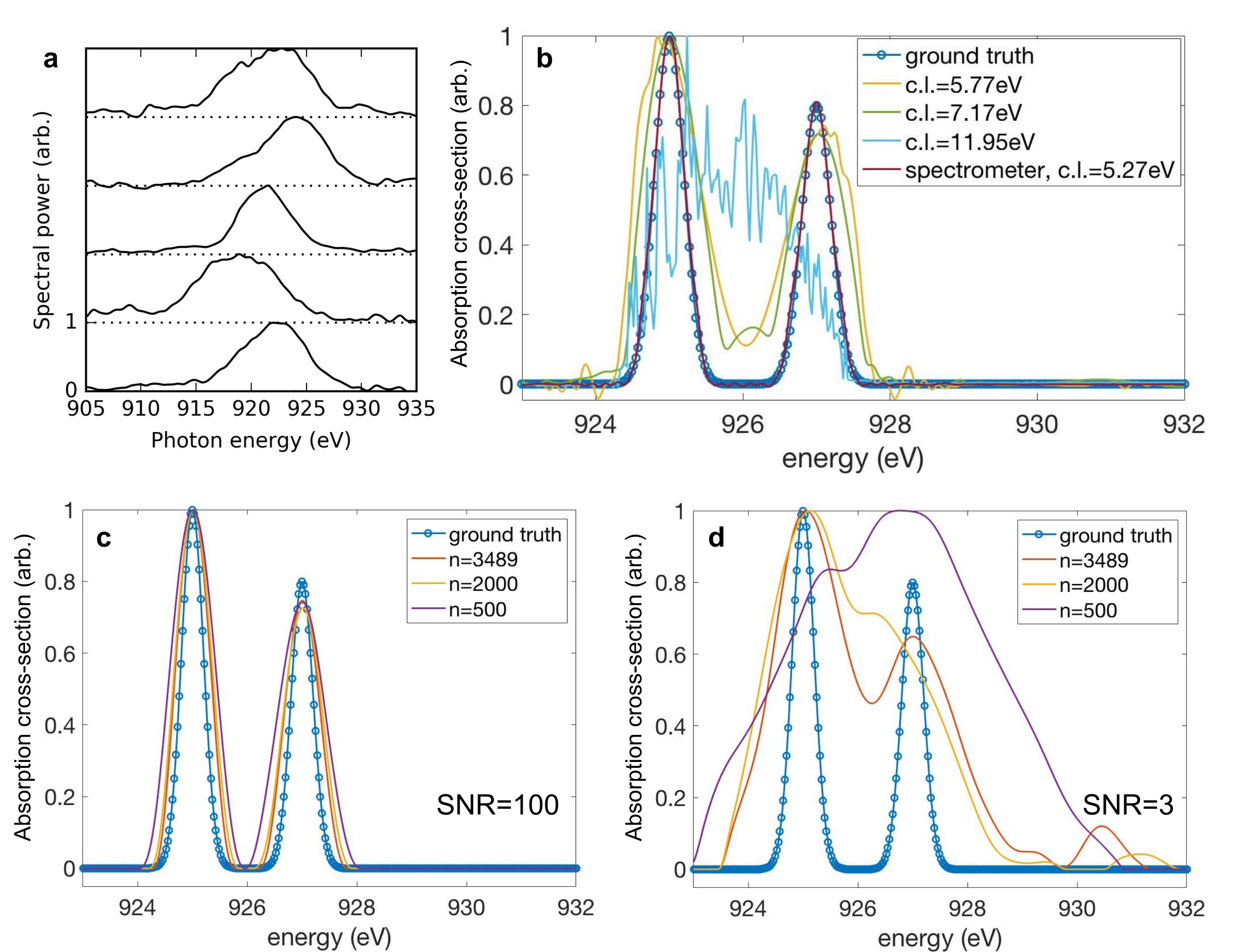}}
\caption{\noindent Performance of ADMM reconstruction. (a)~Single-shot spectra at $\sim$920~eV central photon energy measured with an X-ray spectrometer. (b)~Dependence of resolution of ADMM reconstruction on correlation length~(c.l.) of the photon spectra incident on the sample. The reconstructions are done with 3489 shots. (c)~\&~(d)~Dependence of ADMM performance on number of shots used. Correlation length is kept at 5.27~eV. At a signal-to-noise ratio (SNR) of 100 in the bucket (c), it is possible to faithfully reconstruct the ground truth at a very low number of shots. However, at a lower SNR of 3 (d), a larger number of shots is critical to overcome the noise in the measurement. 
}

\label{fig:mode_corrlen}
\end{centering}
\end{figure}

While the variation of the spectral moments serves as a useful and intuitive guide to assess the variation in our data, fundamentally the resolution is determined by the degree of correlation between neighboring pixels are in the detector. 
This pixel-to-pixel correlation contains information about the inherent variation in the spectra as well as the spectrometer resolution.
%Moreover, it is this variation and the resolution of the X-ray spectral measurement that are the ultimate restricting factors in the resolution of our measurement.
In the limiting case where there is no correlation between spectrometer pixels, the resolution is limited simply by the detector resolution.
As mentioned above, when configured to collect high energy Auger electrons, the resolution of the c-VMI spectrometer (i.e. the measurement of the nitrogen 1s photoelectrons) is significantly diminished compared to what would be available from a typical X-ray photon spectrometer or electron spectrometer under standard operating parameters.
For this reason, in order to further explore the limitations of the correlation absorption spectroscopy technique, we include another set of spectral measurements taken by a photon spectrometer.
The single-shot spectra shown in panel~\textbf{a} of Fig.~\ref{fig:mode_corrlen} are representative examples from a set of 3489 X-ray pulses generated using the ESASE technique at the LCLS~\cite{duris2019}.
To characterize this pixel-to-pixel correlation, we define the correlation length to be the distance between two pixels where the correlation coefficient falls from 1 to $1/e$.
To extract a single correlation length from a set of different spectra, we select the pixels where the averaged spectral intensity is above half maximum, and we average the correlation lengths over the selected pixels.
The correlation length for the measured experimental spectra is 5.27~eV.
To explore the effect of changing correlation length, we start with the experimental data and convolve it with a Gaussian filter of successively increasing width to artificially increase the correlation length of the spectra.
%For this test, we use the higher resolution photon spectrometer measurements.
Again, we simulate a ground truth.
This time we use two Gaussian peaks of rms width 0.2~eV separated by 2~eV, and the simulated bucket is constructed by multiplying the convolved spectra with the ground truth.
%\textbf{reconstruction breaks sooner for worse detector resolution than for less variation. But when detector resolution is infinite, it will be dominated by noise.}
The effect of increasing the correlation length is shown in panel~\textbf{b} of Fig.~\ref{fig:mode_corrlen}.
When the correlation length is below 6~eV, we see that the reconstruction captures the thin width of the ground truth. 
As the correlation length increases, the quality of the reconstruction worsens, as expected.
At a correlation length of 11.95 eV, the reconstruction fails to capture any elements of the spectral shape.% of the 0.2~eV wide peaks separated by 2~eV.

The performance of the matrix inversion is also dependent on the number of shots across which the measurement was taken, particularly in the case of a noisy measurement.
In Fig. \ref{fig:GAspectra6_recon}, we are able to overcome experimental noise by greatly overdetermining our ghost imaging problem.
Analogous to averaging over many shots in a simple spectral measurement, this reduces the overall effect of noise on the measurement.
To explore this limit, we randomly select a subset of the full 3489 spectra and used these to produce simulated bucket measurements, to which we randomly add measurement noise on shot-to-shot basis according to a Gaussian distribution.
We test the sensitivity of the reconstruction using ADMM to the number of shots used in the reconstruction, at low (3:1) and high (100:1) signal-to-noise regimes for the bucket measurement.
Panel~\textbf{c} of Fig.~\ref{fig:mode_corrlen} shows that at high signal-to-noise, it is possible to faithfully reconstruct the absorption spectrum even at low numbers of shots.
In future experiments the counting of electrons or ions could be performed at very high fidelity, with high signal-to-noise and good quantum efficiency.
Panel~\textbf{d} shows that in the scenario where the measurement is noisier, the reconstruction fails to capture the correct width of the absorption features and the quality of the recovered spectrum significantly degrades at a lower number of shots.

%It should be noted that the fundamental unit of resolution for our method is embedded in the resolution of the measurement.
%In particular, the measured correlation length is the critical characterization in that it incorporates the instrument resolution, the variation in the spectrum, and the signal-to-noise in the measurements.
%While the experimental implementation can depend on specific experimental conditions, we emphasize that the correlation length determines the resolution of our technique.

\section*{Conclusions}
In this paper, we have proposed and demonstrated a correlation-based analysis to conduct attosecond transient absorption spectrum (ATAS) measurements at an XFEL facility.
With the development of attosecond pulses from XFELs, attosecond time resolution measurements at an XFEL have become possible.
However, the standard implementation of transient absorption measurements at XFELs will suffer from low spectral resolution due to the wide bandwidth of such short pulses.
The correlation-based technique we discussed in this paper solves this problem by exploiting the inherent shot-to-shot spectral jitter of the X-ray pulses from an XFEL.
We experimentally demonstrated this technique using an experiment where the X-ray spectrum and the photoproduct (i.e. Auger yield) are measured in coincidence, with a single velocity map imaging spectrometer.
This measurement demonstrates excellent agreement with theoretical prediction.
The fundamental limit of the correlation-based analysis is no longer the spectral bandwidth of the X-ray source, but rather the spectrometer resolution and the variation in the X-ray pulse spectrum.
We simulated effects of variation as quantified by the different moments of variation and by the correlation length.
It is worth noting that we found in further simulation that the quality of the reconstruction is more sensitive to decreasing spectrometer resolution than to decreasing variation in the spectra, in the non-noise dominated regime.
Practically, this means that high spectrometer resolution should be prioritized in experimental implementations, without introducing additional measurement noise.
As we have shown, an optimally designed experiment should take into account spectrometer resolution, available variation in the spectra, and the noise level of the measurements.
For retrieving time-resolved absorption spectra the data shots would first have to be sorted on delay, an operation that will be more feasible for the high shot rates anticipated in future high rep-rate machines. 
%The novel technique demonstrated in this paper paves the way for ATAS measurements.

It is possible to extend our method beyond X-ray absorption spectroscopy. 
Provided energy and/or angular resolution in the measurement of the photoproducts, it is possible to reconstruct the photon energy-dependent photoproduct measurement.
Equation~\ref{eq::GIequation} can be easily adapted to include multiple bucket detectors.
In the case of the experimental implementation considered here, this would correspond to a measurement of the energy-dependent resonant Auger spectrum.
Each pixel of the Auger electron kinetic energy spectrum would be an independent bucket measurement.
Then $\mathbf{b}$ becomes a matrix of size $n\times p$, where $n$ is the number of XFEL shots as defined above, and $p$ is the number of pixels in the Auger electron kinetic energy spectrum.
With this change, $\mathbf{x}$ becomes an $m\times p$ matrix, where $m$ is the number of pixels in the absorption spectrum.
This new map, $\mathbf{x}$, is then the photon energy-dependent resonant Auger spectrum.
Such a map is extremely powerful because it provides information not only about the unoccupied electronic orbitals involved in the absorption process, but also on the occupied electronic orbitals that are involved in the Auger process. 
Recording these maps as a function of time-delay would give an even more precise probe of attosecond charge dynamics, similar to the information given in a resonant inelastic X-ray scattering~(RIXS) measurement~\cite{baker2017k}. 

\section*{Acknowledgements}
Use of the Linac Coherent Light Source (LCLS), SLAC National Accelerator Laboratory, is supported by the U.S. Department of Energy~(DOE), Office of Science, Office of Basic Energy Sciences~(BES) under Contract No. DE-AC02-76SF00515.
This work was supported by: DOE-BES Accelerator and detector research program Field Work Proposal 100317; DOE-BES, Chemical Sciences, Geosciences, and Biosciences Division; and  DOE, Laboratory Directed Research and Development program at SLAC National Accelerator Laboratory, under contract DE-AC02-76SF00515. 
P.R. and M.F.K. acknowledge additional support by the DFG via KL-1439/10, and the Max Planck Society.
NB, RO and AL acknowledge support of DOE-BES grant No. DE-SC0012376.

\section*{Author Contributions}
J.P.C., S.L., A.M., and T.D. conceived the research. S.L. and T.D. realized the method. A.M. and J.D. developed the attosecond XFEL source. All authors contributed to collection of the experimental data and writing of the manuscript.

\bibliographystyle{unsrt}

%\bibliography{library,referencesJPC}

\end{document}